\documentclass[singlecol]{epl2}
\usepackage{epsfig,epsf,psfrag}
\usepackage{graphicx}
\usepackage{epic,eepic}
\usepackage{color,pstcol}

\title{Can dephasing generate non-local spin correlations?}

\date{\today}
\author{Colin Benjamin\inst{1}}
\institute{\inst{1} National Institute of Science Education and Research,
IOP Campus, Sachivalaya Marg, Bhubaneswar 751005, India}
\pacs{74.50.+r}{}
\pacs{74.78.Na}{}
\pacs{85.25.-j}{}

\abstract{By examining the full counting statistics of a non adiabatic pure spin pump with particular emphasis on the second and third moments, it is shown that incoherent or sequential transport, in contrast to coherent transport, can change non-local spin shot noise cross-correlations from being anti-correlated to being completely correlated, a truly counterintuitive result. The third moment on the other hand is shown to be much more resilient and its nature remains unaltered in incoherent transport regime. However, phenomenologically including dephasing modifies this picture as both Shot noise and more so the third moment are non-trivially affected. In fact non-local spin correlations are completely positive for maximal dephasing.}

\begin{document}

\maketitle
\section{Introduction} Non-local shot noise correlations, the second moment, in solid state devices have been studied for a long time. Some of these studies include normal metal-superconductor hybrid structures\cite{martin}, coulomb blockaded quantum dots\cite{cottet}, exploiting the Rashba scattering\cite{egues}, etc. However, an experimental demonstration has thus far been lacking. This is mainly due to the difficulty in controlling environmental effects like dephasing or decoherence. The origin of these environmental effects can be traced to magnetic impurities in the experimental system which affects electronic spin and to temperature which leads to electron-phonon interaction and dephasing. It begs the question how to deal with decoherence and reduce it. In this work a novel scheme is proposed in which the dephasing present in such systems can be used as a resource. We particularly concentrate on the electronic spin. The reason for dwelling on the spin instead of charge is because there have been many works on the charge counting statistics however works on the full counting statistics for spin are less visible. However, they have been attempted in different context to that which is the topic of this letter. For example, in \cite{nazarov}, the FCS of spin currents was first attempted, the FCS of spin transfer through ultra small quantum dots in context of Kondo effect was attempted in \cite{komnik} while in \cite{urban} a study of FCS in interacting quantum dots attached to ferromagnetic leads revealed super-poissonian transport. Many works revolve around the non-local spin shot-noise correlations. Among the notable works on non-local spin shot-noise correlations mention may be made of: spin current shot noise of (i) a single quantum dot coupled to an optical micro-cavity and a quantized cavity field\cite{djuric}, (ii) a realistic superconductor-quantum dot entangler\cite{sauret}, and (iii) a spin transistor\cite{he,jian}. In this letter the properties of the third moment are also calculated. The reason for looking at the third moment is because, for charge in contrast to spin transport, the third moment is predicted to be much more resilient to decoherence\cite{jc-been}. In our work we see that this statement does not strictly hold for third moment non-local spin correlations especially when decoherence is large.

 Charge or spin transport is a statistical process involving electrons carrying definite amounts of spin or charge, since charge or spin current fluctuates in time. Therefore, in addition to knowing the mean charge or spin current passing through a normal conductor one needs to know the noise as well as the other transport moments in order to fully characterize charge or spin electron motion. To do this one takes recourse to the full counting statistics(FCS), which gives us the complete knowledge about all the moments of the distribution of the number of transferred charges or spins. The full counting statistics for a non-adiabatic pure spin pump is analyzed in both the completely coherent and incoherent transport  regimes.

In this letter we find that in the coherent transport regime the current and non-local spin shot-noise correlations are similar to that in Ref.\cite{bingdong}. In the sequential or incoherent transport regimes the odd moments are almost unchanged. In contrast the second moment, i.e.,  spin shot noise becomes completely positive. An extremely counter-intuitive result. In absence of any dephasing the third moment spin auto or non-local correlations do not change much from the coherent and incoherent transport regimes. To connect the coherent and incoherent transport regimes we introduce a phenomenological model of dephasing. In this work it is shown that for maximal rate of dephasing the coherent and sequential transport regimes match exactly.

 The main body of this letter starts with an explanation of the model. The coherent density matrix equation is then analyzed separate from the incoherent density matrix equation to bring out the differences. Lastly we bring out a perspective on future endeavors.

\begin{figure}[h]
\centerline{
\includegraphics[width=7cm,height=3cm]{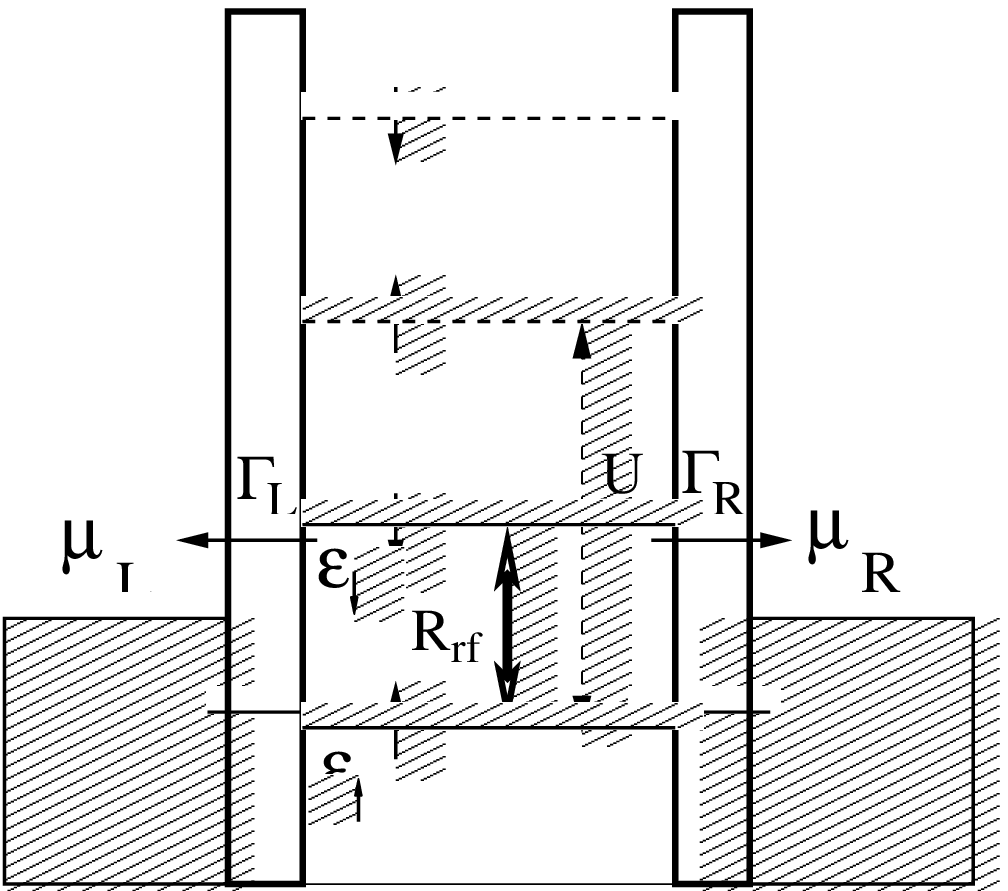}}
\caption{The model system. A single quantum dot is connected to two leads. A large coulomb repulsion is assumed for the quantum dot leading to prevention of double occupation. A large oscillating magnetic field with strength denoted by $R_{rf}=g_{\perp}\mu_{B}B_{rf}/2$ establishes pure spin currents in the leads by pumping spins from lower to higher energy level.}
\end{figure}
\section{Model} The model of Ref.\cite{bingdong} is the
starting point. It is depicted in Fig.1. It is a single quantum dot connected to two leads. The single electron
levels in the dot are split by an external magnetic field  $B$. Thus,
$\epsilon_{\uparrow}-\epsilon_{\downarrow}=g_{z}\mu_{B}B=\Delta$(Zeeman
energy), where $g_z$ is effective electron g-factor in z-direction
and $\mu_B$ is Bohr magneton. No bias voltage is applied across
the leads. An additional oscillating magnetic field
$B_{rf}(t)=(B_{rf}\cos(\omega t),B_{rf}\sin(\omega t))$ applied
perpendicularly to constant field B with frequency $\omega$ nearly
equal to $\Delta$ can pump the electron to higher level where its
spin is flipped, then the spin down electron can tunnel out of the
leads. Coulomb interaction in the quantum dot is considered to be strong
enough to prohibit double occupation. No extra electrons can enter
the quantum dot before the spin-down electron exits. The
Hamiltonian of ESR induced spin battery under consideration is written as:
\begin{eqnarray}
\label{eq_H}
H&=&\sum_{\eta,k,\sigma}\epsilon_{\eta,k,\sigma}c_{\eta,k,\sigma}^{\dagger}c_{\eta,k,\sigma}+
\sum_{\sigma}\epsilon_{\sigma}c_{d\sigma}^{\dagger}c_{d\sigma}+
Un_{d\uparrow}n_{d\downarrow}\nonumber\\
&+&\sum_{\eta,k,\sigma}(V_{\eta}c_{\eta,k,\sigma}^{\dagger}c_{d\sigma}+h.c.)+H_{rf}(t)
\end{eqnarray}
In the above equation,
$c_{\eta,k,\sigma}^{\dagger}(c_{\eta,k,\sigma})$ and
$c_{d\sigma}^{\dagger}(c_{d\sigma})$ are the creation and
annihilation operators for electrons with momentum $k$, spin
$\sigma$ and energy $\epsilon_{\eta,k,\sigma}$ in lead
$\eta(=L,R)$ and for spin $\sigma$ electron on the quantum dot.
The third term describes coulomb interaction among electrons on
the quantum dot. The fourth term describes tunnel coupling between
quantum dot and reservoirs. $H_{rf}(t)$ describes the coupling
between the spin states due to the rotating field $B_{rf}(t)$ and
can be written in rotating wave approximation as:
\begin{equation}
\label{eq_Href}
H_{rf}(t)=R_{rf}(c_{d\uparrow}^{\dagger}c_{d\downarrow}e^{i \omega
t }+ c_{d\downarrow}^{\dagger}c_{d\uparrow}e^{-i \omega t })
\end{equation}
with, ESR rabi frequency $R_{rf}=g_{\perp}\mu_{B}B_{rf}/2$, with
$g$-factor $g_{\perp}$ and amplitude of rf field $B_{rf}$.

The quantum rate equations for the density matrix can be easily
derived as in Ref.\cite{bingdong}. $\rho_{00}$ and $\rho_{\sigma
\sigma}$ describe occupation probability in QD being respectively
unoccupied and spin-$\sigma$ states and off-diagonal term
$\rho_{\uparrow \downarrow (\downarrow \uparrow)}$ denotes
coherent superposition of two coupled spin states in quantum dot.
The doubly occupied is prohibited due to infinite coulomb
interaction $U\rightarrow \infty$. To derive the density matrix,
we proceed as follows. The time dependence can be removed from
Eqs. [\ref{eq_H}-\ref{eq_Href}], by using the following unitary
transformation\cite{zhang}:
\begin{equation}
\label{eq_U}
U=e^{\frac{-i\omega
t}{2}[\sum_{k,\eta}(c_{\eta,k,\downarrow}^{\dagger}c_{\eta,k,\downarrow}^{}-c_{\eta,k,\uparrow}^{\dagger}c_{\eta,k,\uparrow}^{})+(c_{d,\downarrow}^{\dagger}c_{d,\downarrow}^{}-c_{d,\uparrow}^{\dagger}c_{d,\uparrow}^{})]}
\end{equation}
The Hamiltonian is then redefined in the rotating reference form as follows:
\begin{eqnarray}
\label{H_rot} H_{RF}&=&U^{-1}HU+i\frac{dU^{-1}}{dt}U\nonumber\\
&=&\sum_{\eta,k,\sigma}\bar\epsilon_{\eta,k,\sigma}c_{\eta,k,\sigma}^{\dagger}c_{\eta,k,\sigma}+
\sum_{\sigma}\bar\epsilon_{\sigma}c_{d\sigma}^{\dagger}c_{d\sigma}+Un_{d\uparrow}n_{d\downarrow}\nonumber\\
&+&\sum_{\eta,k,\sigma}(V_{\eta}c_{\eta,k,\sigma}^{\dagger}c_{d\sigma}+h.c.)+R_{rf}(c_{d\uparrow}^{\dagger}c_{d\downarrow}+c_{d\downarrow}^{\dagger}c_{d\uparrow})\nonumber\\
\end{eqnarray}
In the above equation,
$\bar\epsilon_{\uparrow}=\epsilon_{D}-\frac{\Delta}{2}+\frac{w}{2}$,
and $\bar\epsilon_{\downarrow}=\epsilon_{D}+\frac{\Delta}{2}-\frac{w}{2}$,
while $\bar\epsilon_{\eta k \uparrow}=\epsilon_{\eta k
}+\frac{w}{2}$ and $\bar\epsilon_{\eta k \downarrow}=\epsilon_{\eta k \downarrow}-\frac{w}{2}$.

 To get the density matrix from the above Hamiltonian, the
following procedure is used. An electron operator affecting only the electron on the dot can be written in terms of $|p><p|, p=0, \uparrow,
\downarrow$. Writing, for the annihilation operator of the
dot $c_{d\sigma}=|0><\sigma|$, and for the creation operator for
the dot $c_{d\sigma}^{\dagger}=|\sigma><0|$, the
Hamiltonian is rewritten in terms of the three states: $|0>, |\uparrow>,
|\downarrow>$, corresponding to empty state, a single electron
with spin-up and single electron with spin-down. The doubly
occupied state in the dot is prohibited by the fact that U is
taken to be extremely large.  {This is the infinite-U Anderson model. We invoke this approximation to avoid double occupancy of the dot. This entails that the probability for an electron entering the dot depends on whether dot is already occupied or not and not on the mean occupation of the dot.} Thus the Hamiltonian reduces to:
\begin{eqnarray}
H&=&\sum_{\eta,k,\sigma}\bar\epsilon_{\eta,k,\sigma}c_{\eta,k,\sigma}^{\dagger}c_{\eta,k,\sigma}+
\sum_{\sigma}\bar\epsilon_{\sigma}|\sigma><\sigma|+Un_{d\uparrow}n_{d\downarrow}\nonumber\\
&+&\sum_{\eta,k,\sigma}(V_{\eta}c_{\eta,k,\sigma}^{\dagger}|0><\sigma|+h.c.)+R_{rf}(|\uparrow><\downarrow|\nonumber\\
&+&|\downarrow><\uparrow|)
\end{eqnarray}
The elements of the density matrix $\rho_{mn}$ in dot spin basis
are expectation values of operators $|n><m|$, with
$n,m=0,\uparrow,\downarrow$, so we can write-$\rho_{00}=<|0><0|>,
\rho_{\sigma\sigma}=<|\sigma><\sigma|>,
\rho_{\sigma\bar\sigma}=<|\bar\sigma><\sigma|> $.
The time evolution of the density matrix elements can be expressed
in terms of expectation values for new operators\cite{thesis_pedersen}. For instance,
\begin{eqnarray}
\label{eq_rho00} i\dot\rho_{00}&=&i \frac{\partial}{\partial
t}<|0><0|>=<[|0><0|,H]>\nonumber\\
\dot\rho_{00}&=&i[H|0><0|-|0><0|H], \nonumber\\
 &=& i[V_{\eta}^{*}|\sigma><0|c_{\eta k \sigma}-V_{\eta}|0><\sigma|c_{\eta k
 \sigma}^{\dagger}],\nonumber\\
&=&[V_{\eta}^{*}G_{\eta k \sigma}^{<}(t,t)-V_{\eta}G_{0 \sigma,
\eta k \sigma }^{<}(t,t)]
\end{eqnarray}

 {To derive the Greens functions for the dot-lead system we assume that the dot-lead system is weakly coupled. This means the dot Greens function in presence of tunneling can be written is the same form as the decoupled dot Greens function. This is also called the Markov approximation in which the probability of a tunneling event at a given time depends only on the occupation at that particular time, i.e. there is no memory structure in the system. Invoking Markov approximation is reasonable as tunneling happens rarely and the system is in same state at each  tunneling event, i.e., when system is weakly coupled.}
The approximated current Green's functions are (using Ref.\cite{thesis_pedersen}) as a guide we have:
\begin{eqnarray}
\label{eq_Gapprx} \small G_{0\sigma,\eta k
\sigma'}^{<}(t,t')&=&\int dt_{1}[G_{0\sigma
\sigma'}^{R}(t,t_{1})V_{\eta k \sigma'}^{*}g_{\eta k
\sigma'}^{<}(t_{1},t')\nonumber\\
&+&G_{0\sigma \sigma'}^{<}(t,t_{1})V_{\eta k
\sigma'}^{*}g_{\eta k \sigma'}^{A}(t_{1},t')],\nonumber\\
G_{\eta k \sigma',0 \sigma}^{<}(t,t')&=&\int dt_{1}[g_{\eta k
\sigma'}^{R}(t,t_{1})V_{\eta k \sigma'}G_{0 \sigma'
\sigma}^{<}(t_{1},t')\nonumber\\
 &+&g_{\eta k
\sigma'}^{<}(t_{1},t')V_{\eta k \sigma'}G_{0 \sigma'
\sigma}^{A}(t_{1},t')]
\end{eqnarray}
The $G_{0\sigma\sigma'}$'s are the green functions for the dot,
while $g_{\eta k \sigma}$ is the Green's function for the
$\eta$-lead in absence of tunneling.

From the convolution theorem for Fourier transforms,
\begin{eqnarray}
\int dt_{1}A(t-t_{1})B(t_{1}-t)=\int du A(u)B(-u)=\int\frac{dw}{2\pi}A(w)B(w).
\end{eqnarray}
Inserting the approximated current Green's functions from Eqs.\ref{eq_Gapprx} into Eq.\ref{eq_rho00} and Fourier transforming one gets:
\begin{eqnarray}
\dot \rho_{00}=|V_{\eta}|^{2}[G_{0\sigma\sigma}^{<}(w)(g_{\eta k
\sigma}^{R}(w)-g_{\eta k \sigma}^{A}(w)) +g_{\eta k
\sigma}^{<}(w)(G_{0 \sigma \sigma}^{A}(w)-G_{0 \sigma
\sigma}^{R}(w))]
\end{eqnarray}

The general property for Green's functions $G^{>}-G^{<}\equiv G^{R}-G^{A}$, is then used-
\begin{eqnarray}
\dot \rho_{00}=|V_{\eta}|^{2}[G_{0\sigma\sigma}^{<}(w)(g_{\eta k
\sigma}^{>}(w)-g_{\eta k \sigma}^{<}(w))+g_{\eta k
\sigma}^{<}(w)(G_{0 \sigma \sigma}^{<}(w)-G_{0 \sigma
\sigma}^{>}(w))]
\end{eqnarray}

The lesser Green's function then becomes-
\begin{eqnarray}
g_{\eta k \sigma}^{<}(t) \equiv <c_{\eta k
\sigma}^{\dagger}c_{\eta k \sigma }(t)>=i e^{-i \epsilon_{\eta k
\sigma}t}<c_{\eta k \sigma}^{\dagger}c_{\eta k \sigma}>
=i e^{-i \epsilon_{\eta k \sigma}t} f_{\eta}(\epsilon_{\eta k
\sigma}),
\end{eqnarray}
where, $f(\epsilon)$ is the Fermi function. Performing a fourier
transformation yields
\begin{eqnarray}
g_{\eta k \sigma}^{<}(w)=2\pi i f_{\eta}(\epsilon_{\eta k
\sigma})\delta(w-\epsilon_{\eta k \sigma}),\mbox{and similarly }
 g_{\eta k \sigma}^{>}(w)=-2\pi i
[1-f_{\eta}(\epsilon_{\eta k \sigma})]\delta(w-\epsilon_{\eta k
\sigma})
\end{eqnarray}
Substituting the above expressions in Eqs.\ref{eq_rho00}, and
using the coupling parameter $\Gamma_{\sigma}^{\eta}(\epsilon)=2\pi\sum_{k}|V_{\eta}|^{2}\delta(\epsilon-\epsilon_{\eta
k \sigma })$ gives-
\begin{equation}
\label{eq_rho_oo_new}
 \dot \rho_{00}=\frac{-i}{2\pi}\int dw\sum_{\eta
\sigma}\{\Gamma_{\sigma}^{\eta}(1-f_{\eta}(w))G_{0\sigma\sigma}^{<}(w)+\Gamma_{\sigma}^{\eta}f_{\eta}(w)G_{0\sigma\sigma}^{>}(w)
\}
\end{equation}
The lesser and greater Greens functions for the dot can be derived
using the same formalism as in Ref.\cite{thesis_pedersen}. Thus,
$G_{0\sigma\sigma}^{<}(w)=2\pi i
\rho_{\sigma\sigma}\delta(w-\epsilon_{\sigma})$, and
$G_{0\sigma\sigma}^{>}(w)=-2\pi i
\rho_{00}\delta(w-\epsilon_{\sigma})$. After substituting these
expressions in Eq.\ref{eq_rho_oo_new}, and integrating gives-
\begin{equation}
\dot \rho_{00}=\sum_{\sigma
\eta}\Gamma_{\sigma}^{\eta}[(1-f_{\eta}(\epsilon_{\sigma}))\rho_{\sigma\sigma}-f_{\eta}(\epsilon_{\sigma})\rho_{00}]
\end{equation}
Now in Ref.\cite{bingdong} the Fermi functions for the left and
right leads with respect to the electron spin
$f_{L}(\epsilon_{\uparrow})=f_{R}(\epsilon_{\uparrow})=1$ and
$f_{L}(\epsilon_{\downarrow})=f_{R}(\epsilon_{\downarrow})=0$.
 {Further the coupling parameters $\Gamma$'s are independent of energy which implies that the density of states and tunneling matrix elements are constant. This approximation is called the wide band limit. It is also assumed that occupations are constant in time as we are only interested in steady state result where this approximation is valid. The weak coupling assumption as invoked above also implies that no broadening of energy level occurs in the dot due to tunneling and this means that coupling is much smaller than temperature.}
Thus,
\begin{equation}
\dot\rho_{00}=-(\Gamma_{\uparrow}^{L}+\Gamma_{\uparrow}^{R})\rho_{00}+(\Gamma_{\downarrow}^{L}+\Gamma_{\downarrow}^{R})\rho_{\downarrow
\downarrow}
\end{equation}
Proceeding in exactly the same way, and using the Ref.\cite{thesis_pedersen} as a guide one can derive the other rate equations as written below. To model incoherence we turn to Ref.\cite{Kieblich} and use that as a model.

\section{Results} We introduce density matrices $\rho_{ab}(t)$
meaning quantum dot is on the electronic state $|a>$
($a=b=0,\uparrow,\downarrow$) or on a  quantum superposition state
$(a\neq b)$ at time $t$. We introduce counting
fields\cite{bagrets}, $\chi_{\eta,\sigma}, \eta=L/R \mbox{ and }
\sigma=\uparrow/\downarrow$ to describe transitions from the dot
to leads.

\subsection{ Coherent transport regime:} We first deal with the coherent regime:
$\dot{\rho}(t)=(\dot {\rho}_{00},\dot {\rho}_{\uparrow \uparrow},
\dot {\rho}_{\downarrow \downarrow},
{\Re(\dot\rho_{\uparrow\downarrow})},
{\Im(\dot\rho_{\uparrow\downarrow})})={\cal M} {\rho}(t)$, with
\begin{equation}
\label{eq_M} {\cal M }=\left (
\begin{array}{ccccc}
-(\Gamma_{L\uparrow}+\Gamma_{R\uparrow})&0&(\Gamma_{L\downarrow}e^{i
\chi_{L\downarrow} }+\Gamma_{R\downarrow}e^{i \chi_{R
\downarrow}})&0&0\\
(\Gamma_{L\uparrow}e^{-i \chi_{L \uparrow}
}+\Gamma_{R\uparrow}e^{-i \chi_{R\uparrow}})&0&0&0&-2R_{rf}\\
0&0&-(\Gamma_{L \downarrow}+\Gamma_{R \downarrow})&0&2R_{rf}\\
0&0&0&-(\Gamma_{L\downarrow}+\Gamma_{R\downarrow})&-\delta_{ESR}\\
0&R_{rf}&-R_{rf}&\delta_{ESR}&-(\Gamma_{L\downarrow}+\Gamma_{R\downarrow})
\end{array}
\right),
\end{equation}
and $\delta_{ESR}=\Delta-\omega$. The normalization relation
$\rho_{00}+\sum_{\sigma \sigma}\rho_{\sigma\sigma}=1$ holds for
the conservation and $\Gamma_{\eta \sigma}=2\pi
\sum_{k}|V_{\eta}|^{2}\delta(w-\epsilon_{\eta k\sigma})$. We
assume the spin relaxation time of an excited spin state into the
thermal equilibrium to be very large.

We calculate the eigenvalues of Eq.\ref{eq_M}. The minimal of
these eigenvalues defines the full counting statistics (as,
$\chi_{\eta \sigma }\rightarrow 0, \eta=L,R;
\sigma=\uparrow,\downarrow$). After finding this eigenvalue
$Ev_{0}$, and then by using the approach pioneered in
Ref.\cite{bagrets}, We calculate the first, second and higher
cumulants. Note that the approach of Ref.\cite{bagrets} has been
generalized in Refs.\cite{Kieblich,sprekeler} to include both
coherent and incoherent transport regimes.
\begin{figure}[h]
\vskip 0.28in
\centerline{\includegraphics[width=8cm,height=5cm]{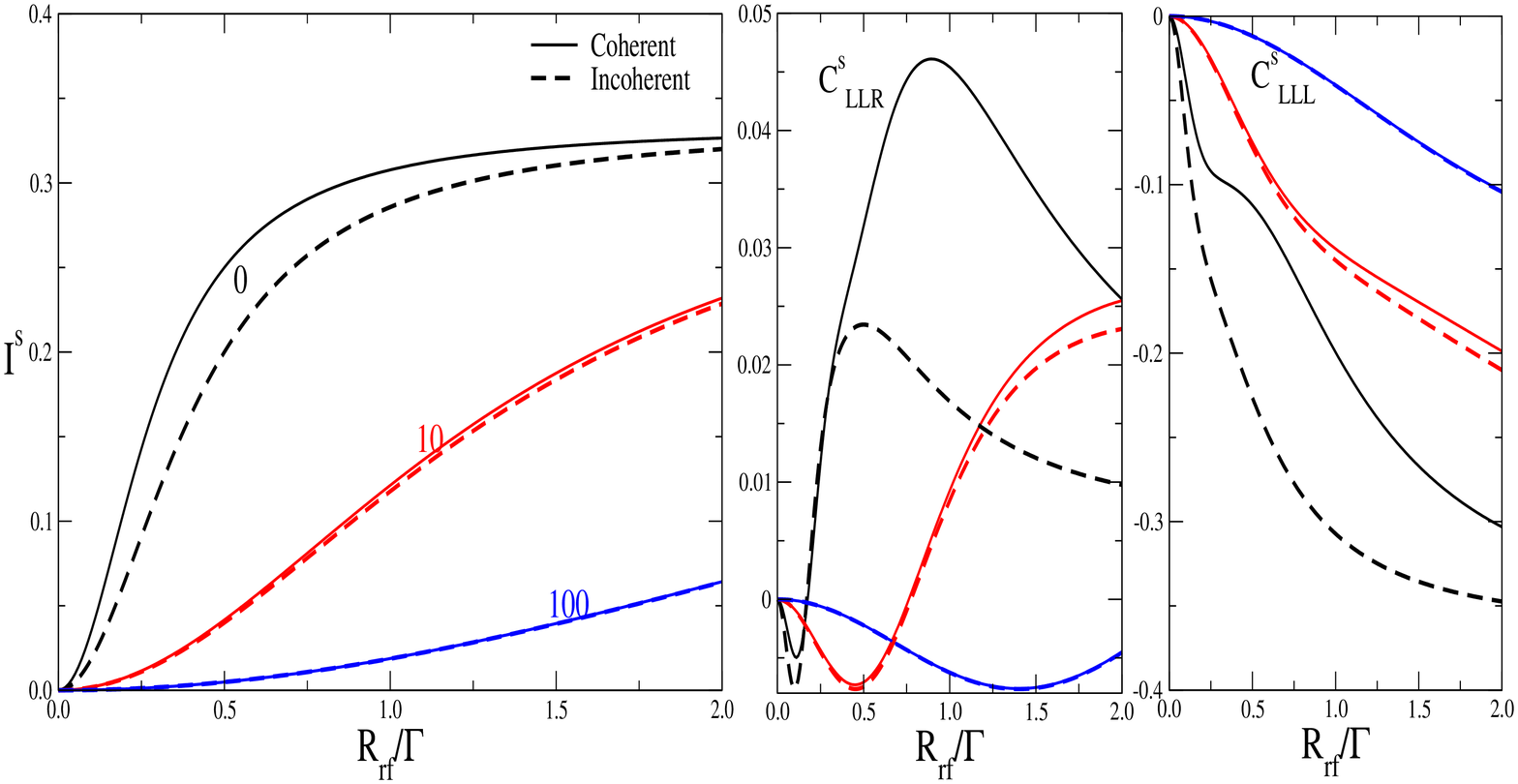}}
\caption{(Color online) A comparison of coherent and incoherent transport
regimes. The odd moments- spin currents(left) and third moment
cross and auto-correlations are plotted as function of strength of rotating field for various dephasing rates $\Gamma_{\phi}/\Gamma=0 \mbox{(black)}, 10 \mbox{(red)}, 100 \mbox{(blue)}$. Solid lines are for Coherent regime and dashed lines are for sequential transport regime. The parameters are $\Gamma=1, \delta_{ESR}=0$.}
\end{figure}

\begin{figure}[h]
\vskip 0.28in
\centerline{\includegraphics[width=8cm,height=5cm]{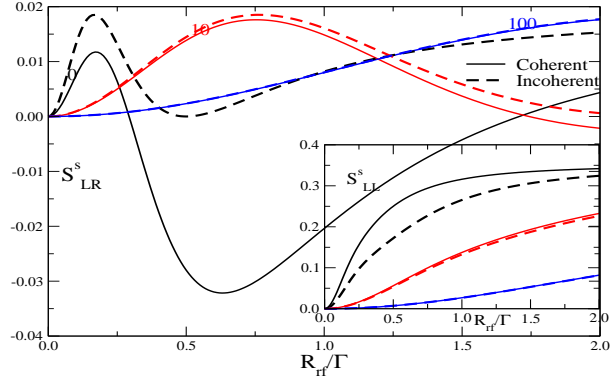}}
\caption{The second moment: The non-local spin shot-noise correlations plotted as function of strength of rotating field for various dephasing rates $\Gamma_{\phi}/\Gamma=0 \mbox{(black)}, 10 \mbox{(red)}, 100 \mbox{(blue)}$. Solid lines are for Coherent regime and dashed lines are for sequential transport regime. In the inset spin auto (or, local) correaltions are depicted. Parameters  are $\Gamma=1, \delta_{ESR}=0$.}
\end{figure}

\begin{figure}[h]
\centerline{\includegraphics[width=8cm,height=8cm]{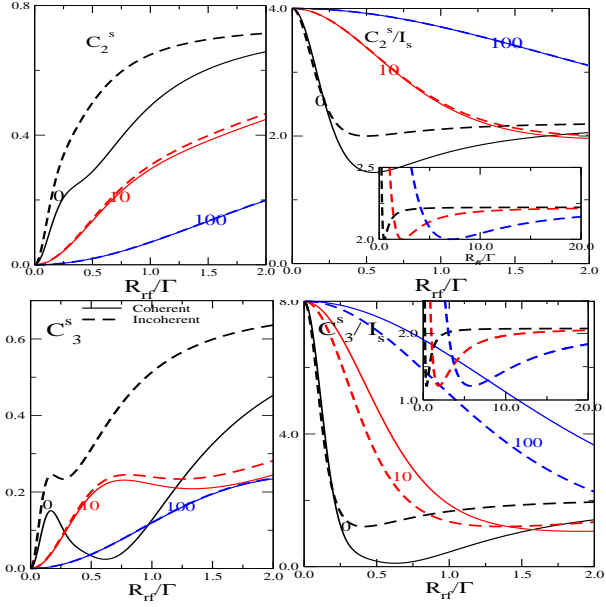}}
\caption{Spin cumulants: $C^{s}_{2}$(top left panel) and $C^{s}_{3}$ (bottom left panel) alongwith their respective Normalized counterparts, Spin Fano factor (top right panel) and Normalized skewness (bottom right panel) plotted as function of strength of rotating field for various dephasing rates $\Gamma_{\phi}/\Gamma=0 \mbox{(black)}, 10 \mbox{(red)}, 100 \mbox{(blue)}$. Solid lines are for Coherent regime and dashed lines are for sequential transport regime, in the insets the large field limit is shown. Parameters  are $\Gamma=1, \delta_{ESR}=0$.}
\end{figure}

The first cumulant is defined as the current, we calculate the
individual spin polarized currents as follows: $I_{\eta
\sigma}=\frac{\partial Ev_{0}}{\partial \chi_{\eta
\sigma}}|_{\chi_{\eta \sigma}\rightarrow 0}$. The spin current is
thus $I^{s}_{\eta}=I_{\eta \uparrow}-I_{\eta \downarrow}$, while
the charge current is $I^{c}_{\eta}=I_{\eta \uparrow}+I_{\eta
\downarrow}$. The second cumulant defines the shot-noise. The shot
noise local and non-local correlations can be calculated as follows.
The spin shot noise local and non-local  correlation is what we
concentrate on.
$S^{s}_{LL}=S^{\uparrow\uparrow}_{LL}+S^{\downarrow\downarrow}_{LL}-S^{\uparrow\downarrow}_{LL}-S^{\downarrow\uparrow}_{LL}$
and
$S^{s}_{LR}=S^{\uparrow\uparrow}_{LR}+S^{\downarrow\downarrow}_{LR}-S^{\uparrow\downarrow}_{LR}-S^{\downarrow\uparrow}_{LR}$
 wherein, $S^{\sigma \sigma'}_{\eta
\eta'}=\frac{\partial^{2}Ev_{0}}{\partial\chi_{\eta\sigma}\partial\chi_{\eta'\sigma'}}|_{\chi_{\eta
\sigma},\chi_{\eta' \sigma'}\rightarrow 0}$. Similarly the third
moment spin correlations are calculated as follows:
$C^{s}_{\eta\eta'\eta''}=C^{\uparrow \uparrow
\uparrow}_{\eta\eta'\eta''}+C^{\uparrow \downarrow
\downarrow}_{\eta\eta'\eta''}+ C^{\downarrow \uparrow
\downarrow}_{\eta\eta'\eta''}+C^{\downarrow \downarrow
\uparrow}_{\eta\eta'\eta''}-( C^{\uparrow \uparrow
\downarrow}_{\eta\eta'\eta''}+C^{\uparrow \downarrow
\uparrow}_{\eta\eta'\eta''} +C^{\downarrow \uparrow
\uparrow}_{\eta\eta'\eta''}+C^{\downarrow \downarrow
\downarrow}_{\eta\eta'\eta''})$, wherein  $C^{\sigma \sigma'
\sigma''}_{\eta
\eta'\eta''}=\frac{\partial^{3}Ev_{0}}{\partial\chi_{\eta\sigma}\partial\chi_{\eta'\sigma'}\partial\chi_{\eta''\sigma''}}|_{\chi_{\eta
\sigma},\chi_{\eta' \sigma'},\chi_{\eta'' \sigma''}\rightarrow
0}$.

The second and third spin cumulants are sum of the individual correlations and are given as $C^{s}_{2}=S^{s}_{LL}+S^{s} _{RR}+2 S^{s} _{LR}$ and $C^{s}_{3}=C^{s}_{LLL}+C^{s} _{RRR}+3 C^{s} _{LLR} +3 C^{s} _{LRR}$. The spin Fano factor and normalized skewness also can provide more information as to how spin transport is affected by dephasing these are defined as- spin Fano factor $C^{s}_{2}/I^{s}$ and normalized spin skewness as $C^{s}_{3}/I^{s}$.

\subsection {Sequential or Incoherent transport regime:}
To go into the incoherent or sequential transport regime as
exemplified in Refs.\cite{Kieblich}, we use the complete coherent
matrix, Eq.\ref{eq_M}, The coefficient matrix for incoherent
transport can be obtained from Eq.\ref{eq_M}, via setting
$\Re(\dot \rho_{\uparrow \downarrow})=0$ and $\Im(\dot
\rho_{\uparrow \downarrow})=0$ and then solving the two
simultaneous equations for $\Re(\rho_{\uparrow \downarrow}) $ and
$\Im(\rho_{\uparrow \downarrow})$ as in
Refs.\cite{Kieblich,sprekeler}. This leads to a $3X3$ matrix:
$\dot{\rho}(t)=(\dot {\rho}_{00},\dot {\rho}_{\uparrow \uparrow},
\dot {\rho}_{\downarrow \downarrow})={\cal M} {\bm \rho}(t)$ with

\begin{equation}
\label{eq_S}
{\cal M }=\left (
\begin{array}[h]{ccc}
-(\Gamma_{L\uparrow}+\Gamma_{R\uparrow})&0&\Gamma_{L\downarrow}e^{i
\chi_{L\downarrow} }+\Gamma_{R\downarrow}e^{i \chi_{R
\downarrow}}\\
\Gamma_{L\uparrow}e^{-i \chi_{L \uparrow}
}+\Gamma_{R\uparrow}e^{-i \chi_{R\uparrow}}&-z&z\\
0&z&-z-(\Gamma_{L \downarrow}+\Gamma_{R \downarrow})
\end{array}
\right),
\end{equation}
and,
$z=\frac{R^{2}_{rf}(\Gamma_{L\downarrow}+\Gamma_{R\downarrow})}{\delta^{2}_{ESR}+(\Gamma_{L\downarrow}+\Gamma_{R\downarrow})^{2}}$.
The minimal eigenvalue of this equation is again what we require.

\subsection{Model for Decoherence}

In order to understand the coherent and sequential transport regimes better and how could the transition between these two regimes be connected we introduce a phenomenological model of decoherence via a charge detector. In this model the off-diagonal elements of Liouville equation are considered to exponentially decay to zero with the rate $\Gamma_\phi$ or $1/T_2$, where $T_2$ is the spin decoherence time, i.e., in the last two rows of the coefficient matrix (\ref{eq_M}), $\Gamma$ is replaced by $\Gamma + \Gamma_\phi$. This method of introducing decoherence can be substantiated by the insertion of a quantum point contact close to the quantum dot. Whenever an electron enters the quantum dot the transmission through it changes. This charge detection leads to exponential damping of the off-diagonals as derived in Refs.\cite{gurvitz,thesis_kiesslich}. Similarly, in the sequential transport regime the decoherence factor can be introduced via the replacement $\Gamma \rightarrow \Gamma + \Gamma_\phi$ in $z$ of the coefficient matrix for sequential tunneling, Eq.(\ref{eq_S}).

In the incoherent regime too the spin current, spin shot-noise auto and cross correlations are calculated and finally the third moment auto and cross-correlations. In Fig. 2, the odd moments are plotted- pure spin current $I^{s}$ and the third moment auto $C^{s}_{LLL}$, and cross-correlations $C^{s}_{LLR}$. In Fig. 3, the second moment, shot noise auto $S^{s}_{LL}$ and cross-correlations $S^{s}_{LR}$. Finally, in Fig. 4 the spin cumulants (both second and third) and the Fano factor and normalized spin skewness are plotted. In the insets of the right hand panels of Fig. 4 the large $R_{rf}$ is shown. One thing which is quite clear in all figures is that dephasing washes out all features in cumulants making them just increase monotonically with increasing strength of the rotating field. Both the spin cumulants are zero for small $R_{rf}$ with increasing $R_{rf}$ the second and third spin cumulants differ markedly, with the third showing much more feature including a pronounced dip at just above $R_{rf}=0.5 \Gamma$. The spin Fano factor plotted on top right hand corner of Fig 4 saturates in the large $R_{rf}$ limit at just above $2$ {\em spin flux quantums} this is in agreement with the normalized skewness which also saturates just under this value.

 In all of these figures  the results for the coherent and incoherent transport regimes are contrasted as dephasing rate is increased from $\Gamma_{\phi}/\Gamma =0$ to $100$. In both regimes the charge current is absolutely zero. Thus there is a pure spin current. The physics behind the pure spin current can be outlined as follows. In the model (Fig. 1) coulomb interaction in the quantum dot is strong enough to prohibit the double occupation, no more electrons can enter the quantum dot before the spin-down electron exits. As a result, the number of electrons exiting from the
quantum dot is equal to that of electrons entering the quantum dot; namely,
the charge currents exactly cancel out each other implying zero charge current.

\section{Conclusions} The pure spin current obtained in our set-up is shown to give rise to completely positive shot noise cross-correlations in the sequential transport regime. This is seen to be sustained when dephasing is included. In fact maximum dephasing gives rise to completely correlated non-local spin correlations. An analysis of the case when there is no dephasing is presented in Table 1. For maximal dephasing the coherent and sequential transport regimes merge.

The main result of our work is depicted in Fig. 3, this is perhaps the first work where it is shown explicitly that the shot noise cross-correlations turn completely positive for maximal dephasing (blue lines in Fig. 3) in either regime. What are the reasons for the completely positive shot noise cross-correlations? One can see from the formula for the spin shot noise cross-correlations it is a difference between same spin and opposite spin correlations. For maximal dephasing one notices that the magnitude of the same spin correlations, which are negative, is always less than that of the opposite spin case. In Fig. 2, the third moment auto and cross-correlations are also plotted. The odd moment doesn't change markedly as  compared to the second moment in case of sequential transport. Dephasing however has a dramatic impact. The third moment auto-correlations are completely negative as expected since the possibility of detecting three electrons is prohibited via Paulli exclusion while third moment cross-correlations turn negative for maximal dephasing. We have compared and contrasted the absolutely incoherent and absolutely coherent regimes. An effective parameter which shows the transition from completely coherent to completely incoherent can be introduced in the coherent density matrix, Eq. 16, to model this. Phenomenologically introducing a spin relaxation time into the coherent density matrix does indeed show the transition between completely coherent and incoherent regimes attesting our results. In this article for the first time the dramatic nature of Non-local spin shot noise correlations as function of dephasing is shown. Future endeavors on effects of incoherence on different geometries especially including superconductors are contemplated.
\begin{table}[h]
\begin{center}
\small
\caption{{Comparing first three moments in coherent and incoherent
(or sequential )transport regimes for zero dephasing}}\vspace{0mm}
\begin{tabular}{|c|c|c|}
\hline
 Moment & {\em Coherent }& {\em Incoherent} \\ \hline
 $1^{st}$ & Pure spin current & Pure spin current \\ \hline
$2^{nd}$ & Non-local shot-noise & Non-local shot-noise\\
&correlations positive for certain&correlations\\
 &range of parameters & always positive.\\
 \hline
 $3^{rd}$ & Third moment finite& Third moment finite\\
 & & {\small No qualitative change} \\
 \hline
\end{tabular}
\end{center}
\end{table}


\begin{thebibliography}{99}
\bibitem{martin} C. Benjamin and J. K. Pachos, Phys. Rev B {\bf 78}, 235403 (2008).
\bibitem{cottet} A. Cottet, W. Belzig, C. Bruder, Phys. Rev. Lett. {\bf 92},
206801 (2004).
\bibitem{egues} J. C. Egues, et. al., Phys. Rev B {\bf 72}, 235326 (2005).
\bibitem{nazarov} A. D. Lorenzo and Y. V. Nazarov, Phys. Rev. Lett. {\bf 93}, 046601 (2004).
\bibitem{komnik} T. L. Schmidt, A. Komnik and A. O. Gogolin, Phys. Rev. B {\bf 76}, 241307(R) (2007).
\bibitem{urban} S. Lindebaum, D. Urban and J. K\"{o}nig, Phys. Rev. B {\bf 79}, 245303 (2009).
\bibitem{djuric}I. Djuric, M. Zivkovic, C. P. Search, G. Recine, arxiv: 0807.2468; I. Djuric and C. P. Search, cond-mat/0611288.
\bibitem{sauret} O. Sauret, T. Martin and D. Feinberg, Phys. Rev. B {\bf 72}, 024544 (2005).
\bibitem{he} Y. He, D. Hou and R. Han, J. Appl. Phys. {\bf 101}, 023710 (2007).
\bibitem{jian}  B. Wang,  J. Wang and H. Guo, Phys. Rev. B 67, 092408 (2003).
\bibitem{jc-been} C. W. J. Beenakker, Journal club for Condensed matter physics October 2003.
\bibitem{bingdong} Bing Dong, H. L. Cui and X. L. Lei, Phys. Rev. Lett. {\bf
94}, 066601 (2005).
\bibitem{zhang} P. Zhang, Q-K Xue and X. C. Xie, Phys. Rev. Lett. {\bf 91},
196602 (2003), Y. Kondo, et.al., quant-ph/0503067.
\bibitem{bagrets} D. A. Bagrets and Y. V. Nazarov, Phys. Rev. B {\bf 67},
085316 (2003).
\bibitem{Kieblich} G. Kie$\beta$lich, P. Samuelsson, A. Wacker and
E. Sch\"{o}ll, Phys. Rev. B {\bf 73}, 033312 (2006).
\bibitem{sprekeler} H. Sprekeler, G. Kie$\beta$lich, A. Wacker and
E. Sch\"{o}ll, Phys. Rev. B  {\bf 69}, 125328 (2004).
\bibitem{thesis_pedersen}  J. N. Pedersen, Cand.scient thesis,
Orsted Laboratory, Niels Bohr Institute fAPG, University of
Copenhagen, June 2004.
\bibitem{gurvitz} S. A. Gurvitz, Phys. Rev. B {\bf 56}, 15215 (1997).
\bibitem{thesis_kiesslich} G. Kiesslich, Ph. D thesis, TU Berlin, October 2005.
\end{thebibliography}
\end{document}